\documentclass[aps,prc,twocolumn,superscriptaddress]{revtex4}

\usepackage{longtable}
\usepackage{graphicx}
\usepackage{epsfig}
\usepackage{dcolumn}
\usepackage{bm}
\usepackage{multirow}
\usepackage{color}
\usepackage{sidecap}
\usepackage[normalem]{ulem}

\begin{document}

\title{Excited states of the odd-odd nucleus $^{158}$Eu from the (d,$\alpha$) reaction}

\author{D. Bucurescu}
\affiliation{ Horia Hulubei National Institute of Physics and Nuclear Engineering,
P.O. Box MG-6, R-76900 Bucharest, Romania}

\author{S. Pascu}
\affiliation{ Horia Hulubei National Institute of Physics and Nuclear Engineering,
P.O. Box MG-6, R-76900 Bucharest, Romania}

\author{T. Faestermann}
\affiliation{Physik Department, Technische Universit\" at M\" unchen,
D-85748 Garching, Germany}

\author{H-F. Wirth}
\affiliation{Fakult\" at f\" ur  Physik, Ludwig Maximilians Universit\" at M\" unchen, Am
Coulombwall 1, 85748 Garching, Germany}

\author{C. Costache}
\affiliation{ Horia Hulubei National Institute of Physics and Nuclear Engineering,
P.O. Box MG-6, R-76900 Bucharest, Romania}

\author{A. Ionescu}
\affiliation{ Horia Hulubei National Institute of Physics and Nuclear Engineering,
P.O. Box MG-6, R-76900 Bucharest, Romania}
\affiliation{Faculty of Physics, University of Bucharest, 405 Atomi\c stilor, Bucharest - M\u agurele,
Romania}

\author{R. Lic\u a}
\affiliation{ Horia Hulubei National Institute of Physics and Nuclear Engineering,
P.O. Box MG-6, R-76900 Bucharest, Romania}

\author{R. Mihai}
\affiliation{ Horia Hulubei National Institute of Physics and Nuclear Engineering,
P.O. Box MG-6, R-76900 Bucharest, Romania}

\author{A. Turturic\u a}
\affiliation{ Horia Hulubei National Institute of Physics and Nuclear Engineering,
P.O. Box MG-6, R-76900 Bucharest, Romania}

\author{R. Hertenberger}
\affiliation{Fakult\" at f\" ur  Physik, Ludwig Maximilians Universit\" at M\" unchen, Am
Coulombwall 1, 85748 Garching, Germany}

\date{\today}


\begin{abstract}

Excited states in the $^{158}$Eu nucleus  have been determined with the 
 $^{160}$Gd(d,$\alpha$)$^{158}$Eu reaction, studied at an incident energy of
18.0 MeV with the Munich tandem and Q3D spectrograph. More than 50 excited 
states have been determined up to 1.6 MeV excitation, some of them corresponding 
to states previously observed in the $\beta^-$-decay of $^{158}$Sm.
The number of levels found in this nucleus at low excitation energies 
follows the systematic trend of the level densities in the other isotopes with mass 
152--156.
\end{abstract}


\maketitle


\section{Introduction}

The study of nuclear structure in  rare earth nuclei with a 
multitude of nuclear reactions has been rather intensive
especially in the  region near the neutron number $N = 90$ where the nuclear properties 
undergo a rapid change,  pinpointing one of the best examples  of quantum shape phase 
transition. While the even-even nuclei and odd-mass nuclei are relatively well studied, 
the odd-odd nuclei in this region are less investigated. One of the possible study tools, 
making use of the many available stable targets in this region, is the (d,$\alpha$) reaction. 
When performed on even-even targets, it leads to odd-odd nuclei, and the advantage is that 
the target  has a $0^+$ ground-state, which facilitates the determination of the spin and 
parity of the states in the odd-odd nucleus. Rather surprisingly, this powerful tool 
was practically unused in the rare earth nuclei. With the exception of the reaction 
$^{152}$Sm(d,$\alpha$)$^{150}$Pm [and of two other reactions used for its energy calibration, 
$^{140}$Ce(d,$\alpha$)$^{138}$La  and $^{142}$Nd(d,$\alpha$)$^{140}$Pr], which was 
used to determine the level structure of the practically 
unknown $^{150}$Pm nucleus \cite{150pm}, this reaction was never performed on other targets in 
the rare-earth region. 

We decided to use this reaction in order to determine the level structure of the $^{158}$Eu 
odd-odd nucleus. For this nucleus there are no adopted levels in the ENSDF database \cite{ENSDF}, except 
for a ground state with a proposed spin-parity (1$^-$) as expected from Nilsson configurations. 
The ENSDF evaluation mentions, however, determinations of excited levels of $^{158}$Eu in 
an unpublished study of this nucleus
by the ${\beta}^-$-decay of $^{158}$Sm, which were also used in a publication where an 
analysis of the total absorption $\gamma$-spectrum in the $\beta$-decay 
was performed \cite{beta158Sm}.

The study of the (d,$\alpha$) reaction on chains of even-even targets, such as that of Nd, 
Sm, Gd, and Dy nuclei, would be of considerable interest also because it may offer a systematic 
view of the structure evolution of the odd-odd nuclei, an aspect that will be exemplified
at the end of this work.

\section{Experiment and results}
The experiment was performed at the Munich tandem accelerator, using a deuteron beam of 18 MeV and 
a 0.5 $\mu$A average intensity. 
The target was 
125 $\mu$g/cm$^2$  Gd$_2$O$_3$ 98.2$\%$ enriched in $^{160}$Gd on 10 $\mu$g/cm$^2$ Carbon foil.
Its main impurities were $^{158}$Gd, $^{157}$Gd, and $^{156}$Gd, each less than 1$\%$. 
The reaction products were analyzed in the Q3D spectrograph \cite{Q3D} and detected and identified 
in its focal plane detector, a multiwire proportional chamber with readout of a cathode 
with microstrip foil structure for ${\Delta}E-E$ particle identification and position determination \cite{focal}.  

Spectra were recorded at an angle of 10$^\circ$ relative to the beam direction, with an acceptance of the 
spectrograph of 14.61 msr ($21.8 \times 24.5$ mm$^2$). 
Figure 1 displays a  ${\Delta}E-E$ plot for the reaction products that enter the focal plane detector,
showing the good separation of the $\alpha$-particles. 
The other events from this plot very likely represent tritons, deuterons, and $^3$He (from left to right), although a sure identification is difficult due to the different reaction $Q$-values, extended range of energies of the emergent particles, the rather compressed scale of the rest energy axis, and the proximity to the threshold cutoff.
With this identification of the 
$alpha$'s the spectra of the $(d,\alpha)$ reaction were practically background-free.
The beam current was integrated into a Faraday cup placed after the target in order to
determine the cross sections.

Due to the small cross-sections of our reaction and the available beam intensity and 
measurement time, angular distributions could not be measured. We concentrated on the measurement 
at just one angle, of 10$^\circ$. Figure 2(a) shows the 10$^\circ$ spectrum measured during a total of 
19 hours. The energy calibration of this spectrum has been achieved by measuring, in the same 
conditions, the spectrum of the $^{111}$Cd(d,$\alpha$)$^{109}$Ag reaction, with a target of
150 $\mu$g/cm$^2$ thickness, for which peaks 
corresponding to well-known levels of $^{109}$Ag \cite{ensdf109Ag} have been identified.  This calibration spectrum is shown 
in Fig. 2(b). 
\begin{figure*}
\vspace*{9mm}
\epsfig{figure=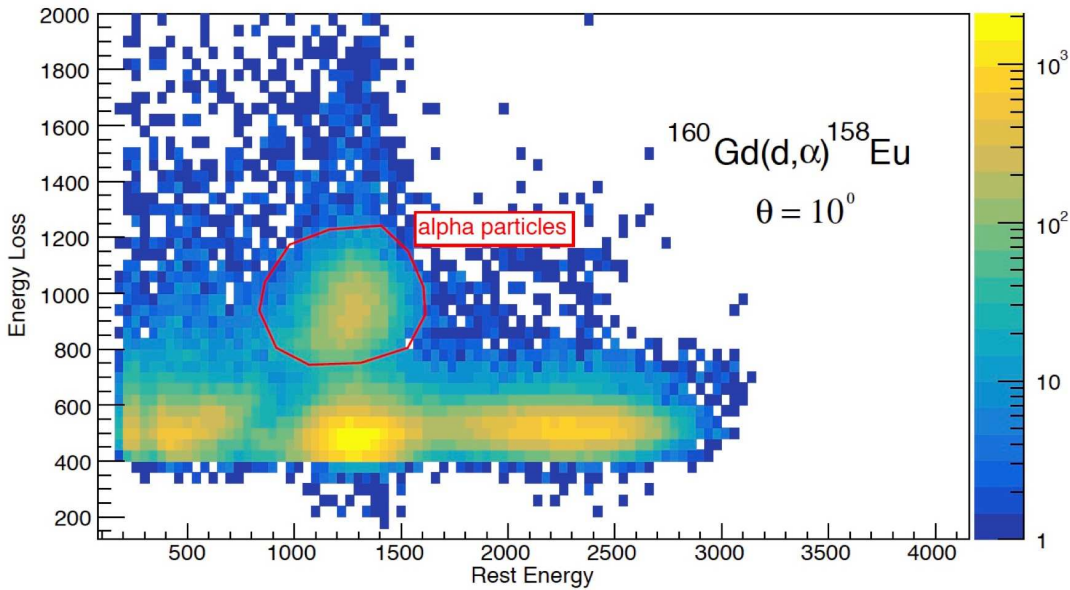,width=0.8\textwidth,angle=0}
\caption{(Color online) Graph of the energy loss versus the rest energy (both in arbitrary units)
of the reaction products that reach the focal plane detector, showing the good separation
of the alpha particles. } 
\label{fig 1}
\end{figure*}

\begin{figure*}
\vspace*{9mm}
\epsfig{figure=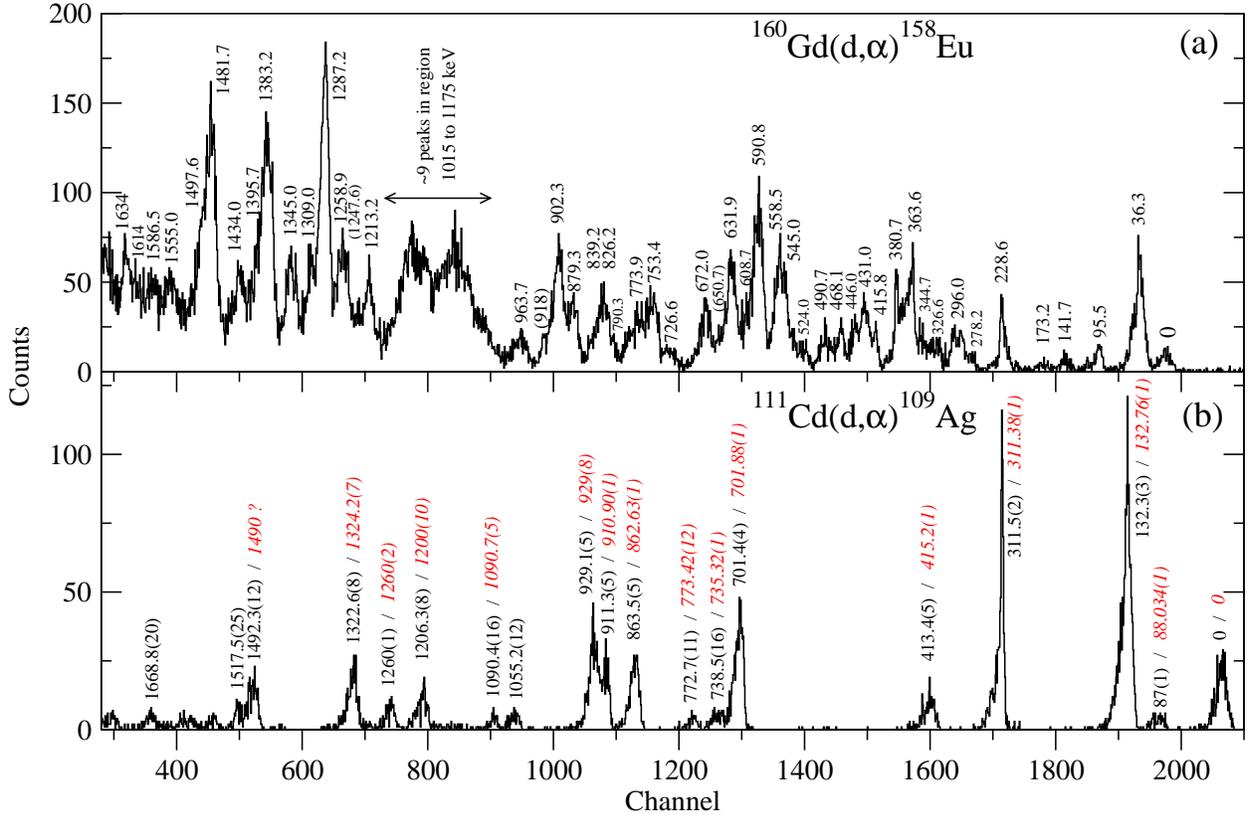,width=0.6\textwidth,angle=270}
\caption{(Color online) Spectra measured at 10$^\circ$ with the same magnetic settings of the spectrograph for
(a) our reaction, and (b) the reaction used for energy calibration, 
$^{111}$Cd(d,$\alpha$)$^{109}$Ag. In spectrum (b) the peaks are labeled both with 
the ENSDF adopted energies \cite{ensdf109Ag} (in red italics) and those assigned
with the calibration curve, respectively. In spectrum (a) the peaks are labeled 
with the excitation energies of the states in $^{158}$Eu, as  
found with the calibration curve (see text and Table I). The spectrum in (a) was obtained in 
19 h of measurement with a beam of about 0.5 ${\mu}A$. For comparison, 
the spectrum in (b) was produced in 100 min under similar conditions. } 
\label{fig 2}
\end{figure*}

Both spectra in Fig. 2 have been processed with the GASPAN peak fitting program \cite{gaspan}.
The FWHM energy resolution was about 15 keV for the spectrum in Fig. 2(a) and 12 keV for that 
in Fig. 2(b), respectively. 
Peaks due to the target impurities were not visible in the spectrum 
of Fig. 2(a).
For the calibration spectrum in Fig. 2(b), an energy calibration curve 
for the excitation energy $E_x$ in $^{109}$Ag versus channel 
number was generated as a 
second degree polynomial. From the peak energy 
labels in Fig. 2(b) one can see that this curve describes the excitation energies known 
with good precision \cite{ensdf109Ag} with an accuracy of less than 1.5 keV. 
This calibration curve was then transformed, 
by kinematics calculations, in a new calibration curve $E_{abs}$ versus channel number, where
$E_{abs}$ is the absolute energy of the $\alpha$-particles (of the order of 27 MeV). 
This second calibration curve was used for the spectrum in Fig. 2(a) in order to determine the
absolute $\alpha$-particle energies of the peaks corresponding to states in $^{158}$Eu, which 
were then transformed into excitation energies by using kinematic calculations. This procedure
was necessary in order to take into account the rather different recoil energies of the 
residual nuclei in the two reactions, due to the large mass difference between the target nuclei.  

{\it Q-value of the} $^{160}$Gd(d,$\alpha$)$^{158}$Eu {\it reaction}. A better determination of
this quantity resulted as a byproduct of the energy calibration described above. The $Q$-value of
the calibration reaction (on the $^{111}$Cd target) is rather well known, 
$Q_{(d,\alpha)}$($^{111}$Cd) = 10178.0$\pm$1.3 keV, 
as given in the 2016 mass table \cite{mass2016}. For the reaction  $^{160}$Gd(d,$\alpha$)$^{158}$Eu
the $Q$-value is given as $Q_{(d,\alpha)}$($^{160}$Gd) = 10024$\pm$10 keV \cite{mass2016}. By using the $Q$-value of the $^{111}$Cd target, our
measurement of the energy of the peak corresponding to the ground state of $^{158}$Eu (Fig. 2(a))
provided a  value of Q$_{(d,\alpha)}$($^{160}$Gd) = 10035.5$\pm$1.6 keV, 
which is consistent with the older value but more precise.

{\it Excited states of the} $^{158}$Eu {\it nucleus}. 
Table I shows the energy levels found for $^{158}$Eu in the present experiment, 
 In both Table I and Fig. 2 the errors given for the energy values are the statistical
errors, as resulted from the calibration curve and the errors in the peak centroids. As one can 
see from Fig. 2, the calibration curve (second degree polynomial) 
deduced from the reaction on the $^{111}$Cd target works well up to an excitation energy of  
1.32 MeV, corresponding to an excitation energy in $^{158}$Eu of about 1.23 MeV. Beyond this 
excitation energy, up to the highest excited state determined (about 1.6 MeV) the energies 
given in the table are based on the
extrapolation of the calibration curve.  It is therefore expected that
with increasing energy this procedure may provide increasing deviations from the (unknown) real 
energies, that are larger than the specified statistical error.
Also, to better see the basis of the peak assignments, Fig. 3 shows details of the peak 
fitting with the GASPAN program. The peak shapes were fitted with a gaussian plus a left side
(lower $\alpha$-particle energy) exponential tail which is due to the energy loss of the alpha particles in the thin target. 
A fixed tail fraction was chosen, which was found by eliminating the tendency to fit the peaks as doublets, and by a good description of the shape of strong, better separated peaks.
Figure 3 shows six panels corresponding to fits in the 
six adjacent regions of
the total spectrum shown in Fig. 2(a). 
Some weaker fits, e.g., those to the 95.5 keV and 228.6 keV peaks may be due to the fact that their shape did not reach stability yet due to the weak statistics. Attempts to fit the 228.6 keV peak by a doublet failed, while for the 95.5 keV peak such a procedure was not justified due to the low number of counts.
Tentative levels (shown within parentheses in Table I) correspond to rather small, less certain peaks 
found through the peak decomposition procedure. Figure 3(e)
corresponds to the region between 1015 and 1175 keV excitation, where there
are states with significant overlap (average spacing comparable with the energy resolution).
The peak decomposition from this region should be considered with some caution. 
The number of states found by GASPAN in this region depends somewhat on the width allowed 
for the peaks; by imposing a FWHM value comparable to that in the adjacent regions (with 
better separated peaks) one finds a number of nine peaks in this region, two of them being 
tentative (see Table I). The other (stronger) peaks found in this region appear to be 
relatively stable to reasonable variations allowed for the widths of the peaks.

As a result of the analysis of the (d,$\alpha$) spectrum of Fig. 2(a), a number of 
58 excited states have been assigned in the $^{158}$Eu nucleus (five of these being tentative)
up to about 1.6 MeV excitation. 
In Table I they are compared with the 27 excited states proposed from the $\beta^-$-decay of 
$^{158}$Sm in the same energy range \cite{ENSDF,beta158Sm}. 
Fourteen of these states may coincide  with states observed in the $\beta$-decay experiment. 
\newline

\LTcapwidth=8.7cm
\begin{longtable}{ccc}
\caption{\label{Table III} Energy levels of $^{158}$Eu as observed in the present 
(d,$\alpha$) reaction experiment, compared to levels observed in the 
$\beta^-$-decay study of $^{158}$Sm \cite{ENSDF,beta158Sm}. 
When the energies of 
levels from the two experiments differ by less than 3 keV, they  
are placed on the same line and it is assumed that they may represent the same excited state. 
Levels tentatively proposed in our experiment 
are given within parentheses (see also Fig. 3). 
The groups labeled by
(a), (b), ... , (f) correspond to the six graphs in Fig. 3. }\\
\\
\multicolumn{2}{c}{Present experiment}&{$\beta$-decay \cite{ENSDF,beta158Sm}}\\
 E$_x$ (keV)  & $\frac{d\sigma}{d\Omega}$(10$^\circ$) [$\mu$b/sr] & E$_x$ (keV)\\
\hline
		& group (a) \\
0		&  0.09	&   0 	\\
36.3(7) 	&  0.37	&38.9 	\\
95.5(10)	&  0.06 & 97.7 	\\
141.7(12)	&  0.04 \\
173.2(16)	&  0.03  	\\
		& 	&189.5	\\
\hline
		& group (b)\\
		&	& 224.2 \\
228.6(8)	& 0.17	& 229.9	\\
278.2(12)	& 0.07	& 	\\
296.0(9)	& 0.15 	& 295.8	\\ 
326.6(11)	& 0.10	& 324.7	\\
		& 	& 338.8 \\
344.7(13)	& 0.10	&       \\
363.6(8)	& 0.34	& 363.6 \\
		&	& 373.4	\\
380.7(9)	& 0.20	\\
\hline
		& group (c) \\
415.8(15)	& 0.12	&\\
431.0(10)	& 0.22	&\\
446.0(14)	& 0.10	&\\
468.1(10)	& 0.11	& 467.8 \\
		&	& 470 \\
490.7(10)	& 0.10	&\\
		&	& 507.3 \\
524.0(13)	& 0.08	& \\
545.0(15)	& 0.16	& \\
		&	& 551.3 \\
558.5(8)	& 0.41	& \\
590.8(7)	& 0.63	& \\
608.7(16)	& 0.11	& \\
631.9(8)	& 0.37	& 632.8 \\
$[650.7(26)]$	& 0.05	& \\
		&	& 660 \\
672.0(8)	& 0.21	& \\
\hline	
		& group (d) \\
726.6(13)	& 0.08	& \\
		&	& 741.1 \\
753.4(8)	& 0.29	& \\
773.9(14)	& 0.16	& \\
790.3(18)	& 0.08	& 791.5 \\
826.2(12)	& 0.20	&  \\
839.2(16)	& 0.15	& \\
879.3(8)	& 0.24	& \\
902.3(8)	& 0.43	& \\
$[918(5)]$	& 0.04	& 921.3 \\
\hline
		& group (e) \\
963.7(15)	& 0.10	& \\
$[1016(3)]$	& 0.10  & 1010 \\
1032.2(17)	& 0.16	& \\
1052.1(10)	& 0.35	& \\  
1072.3(10)      & 0.43  & \\
1093.2(10)	& 0.31	& \\
		&	& 1110 \\
1118.7(9)	& 0.37  & \\
1139.9(13)	& 0.38	& \\
1155.0(19)	& 0.22	& \\
$[1174.3(28)]$	& 0.06	& \\   
\hline
		& group (f) \\
		& 	& 1209.6 \\
1213.2(13)	& 0.24	& \\
$[1247.6(22)]$	& 0.14	& \\
1258.9(15)	& 0.30	& \\
1287.2(12)	& 1.01	& \\
1309.0(34)	& 0.06	& \\
1345.0(15)	& 0.23	& 1342.9 \\
1383.2(16)	& 0.70	& \\
1395.7(19)	& 0.31	& 1395.3 \\
		& 	& 1421.0 \\
1434.0(19)	& 0.18	& \\
		&	& 1448.0 \\
1481.7(19)	& 0.84	& \\
1497.6(22)	& 0.31	& \\
\hline
1555.0(25)	& 0.17  & 1550 \\	
1586.5(26)	& 0.15	\\
1614(3)		& 0.13	\\
1634(3)		& 0.25	\\
\hline
\end{longtable}

\begin{figure*}
\vspace*{9mm}
\epsfig{figure=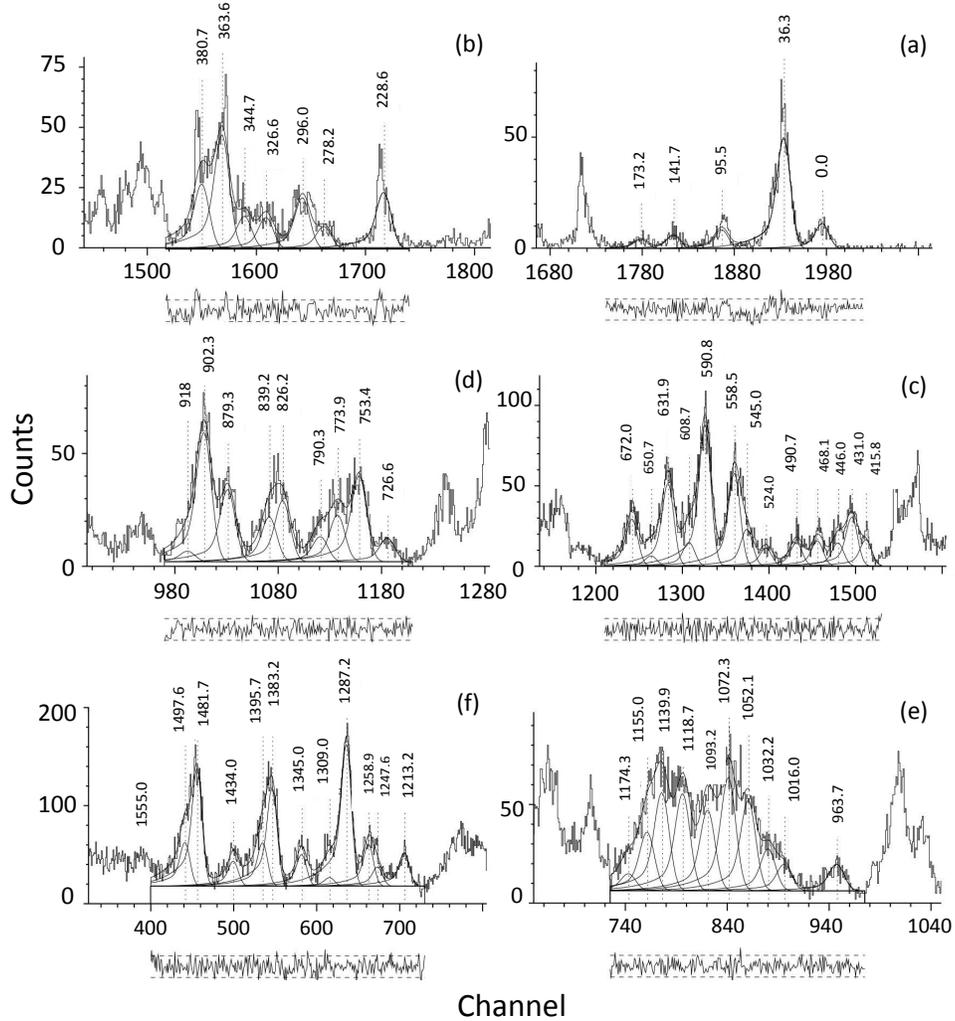,width=0.7\textwidth,angle=0}
\caption{The GASPAN program fits to the spectrum of Fig. 1(a). The peaks are labeled 
with the level energy in keV (see Table I). A residue spectrum with two standard deviations
statistical limit is shown below each graph. The six graphs 
correspond to the six groups of levels displayed in Table I.} 
\label{fig 3}
\end{figure*}

\section{Discussion and conclusions} 

As a result of the present experiment and of the unpublished $\beta$-decay 
study \cite{ENSDF,beta158Sm} a large number of excited states of $^{158}$Eu 
have been determined up to about 1.5 MeV excitation (Table I). 
No spin and parity
values were assigned to any of these levels. In the $\beta^-$ decay of the 
$0^+$ ground state of $^{158}$Sm the populated states are expected to have spin values 
0, 1, and 2$\hbar$ and many of these were populated in our reaction too. The spin window of the states 
seen in the (d,$\alpha$) reaction is wider, states up to spin 
about 6 $\hbar$ may be populated (see, e.g., Ref. \cite{196Au}), with higher spin states being favored due to the large angular momentum mismatch of the reaction. 

Although without spin and parity value assignments, the knowledge
available now on this
odd-odd nucleus extends the nuclear structure systematics of 
the odd-odd Eu isotope chain, and allows a new, stimulating view of 
this interesting mass region. 

It was recently shown that the nuclear level density can be employed as a useful
indicator of the  critical shape phase transitions (SPT) in nuclei
\cite{QPM}. The 
connection between the evolution of the level density at low excitation 
energies and the phase transition phenomenon was examined in detail in the 
rare earths region, where there is the well known first order SPT  that
takes place around the critical point $N \approx 90$. This behavior is 
induced by the variation of a non-thermal control parameter -- the number of
neutrons $N$. The SPT manifests itself by  
a rapid evolution of the ground-state equilibrium deformation around the 
critical point, which is reflected in discontinuous variations of different 
so-called effective order parameters (such as the two-neutron binding 
energy, nuclear radii, etc.) as a function of $N$. 
The level density was shown to display a maximum value
at the critical point \cite{QPM}, which is also consistent with  the phenomenon of 
phase coexistence in nuclei at, or close to, the critical point.  

The critical shape phase transitions were less studied in the odd-odd 
nuclei. Experimental determinations of the level density are rather scarce in 
such nuclei. In particular, the only isotopic chain for which systematic 
data exist is that of the Europium  \cite{QPM}. 
Experimental level densities at low excitation energies were taken from 
Ref. \cite{BSFG}, where the parameters of simple level density models, such as
the back-shifted Fermi gas (BSFG) or the constant temperature (CT) models 
were determined by fitting the experimental low-excitation {\it complete} level 
schemes and the level density at the neutron binding energy. 
In the BSFG model, the total level density is described as
$\rho(E) = \frac{e^{2\sqrt{a(E-E_1)}}}{12\sqrt{2}\sigma a^{1/4}(E-E_1)^{5/4}}$, where 
$E$ is the excitation energy,  $a$ and 
$E_1$ are two empirical parameters and $\sigma$ is the spin cutoff parameter
\cite{BSFG}.  
The parameter $a$ of the BSFG model may be taken as a 
measure of the level density: for nuclei with comparable masses, the larger $a$, 
the larger is the level density \cite{QPM}. 
Figure 4 shows the evolution of the experimental $a$ parameter known for
three odd-odd Eu isotopes: $^{156}$Eu,   $^{154}$Eu, and $^{152}$Eu.
For these three isotopes, the knowledge of the low-excitation level scheme is
considered complete within the following excitation energy/spin windows:
(0--0.39 MeV)/(0 -- 5$\hbar$) for $^{156}$Eu, (0 -- 0.49 MeV)/(1 -- 5$\hbar$) for $^{154}$Eu, 
and (0 -- 0.35 MeV)/(1 -- 4$\hbar$) for $^{152}$Eu, respectively \cite{BSFG}. In Fig. 4 
it is seen that the experimental $a$ has the largest value at $N=89$, 
near the critical point of the control parameter $N$, and decreases with increasing $N$. 

Since $^{158}$Eu is  far from the critical point of the phase transition, we expect a 
relatively low level density in this nucleus, compared to that of the isotopes of mass 152 to 156. 
In order to examine the available data from a larger $N$ region  
we adopt here a simplified procedure.
 For this, we will directly compare the number of  levels
known in these nuclei up to an excitation energy of 0.35 MeV. 
This excitation energy range was chosen because it is common 
to the three nuclei in which 
complete level schemes exist ($N =$ 89, 91, and 93). The 
number of states up to 0.35 MeV is 83, 60, and 24 for $N=$ 89, 91, and 93, respectively \cite{ensdf-Eu}.

\begin{figure}[tbp]
  \begin{center}
    \includegraphics[height=8.0cm,angle=270]{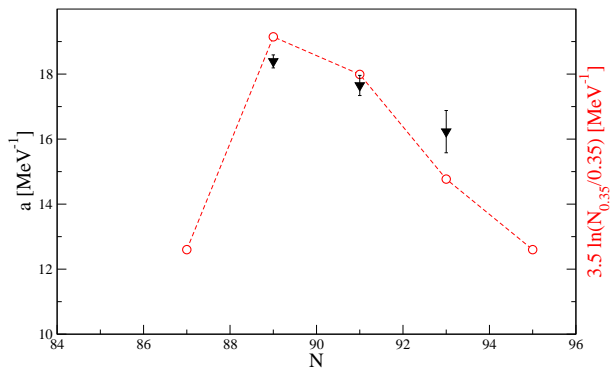}
  \end{center}
  \caption{(Color online) The experimental $a$ parameter of the BSFG model level densities \cite{BSFG}
(black triangles) and the simplified level density of levels up to 0.35 MeV 
excitation (circles and dotted line). 
}
	{\label{fig 4}}
\end{figure}

For $N=95$ ($^{158}$Eu) we count a number of 13 levels up to 0.35 MeV excitation
(Table I). 
Given the spin values covered by the two experiments , it is likely that 
this level scheme is well known up to this energy, close to 
completeness (within the same spin range as that of the three lighter isotopes).
Actually, a few 
missing levels would not significantly alter our conclusions. 
For $N=87$ ($^{150}$Eu), we have a similar situation, with a number 
 of about 13 levels \cite{ensdf-Eu}.
In Fig. 4 we represent also a rough level density determined as 
the number of levels per MeV, 
$N_{0.35}$/0.35 (where $N_{0.35}$ is the number of levels counted up to 
an excitation energy of 0.35 MeV), arbitrarily normalized such as its logarithm approximately scales
as the $a$ parameter.
This approximate low-energy level density shows the same pattern as that of the
experimental $a$ parameter. $^{158}$Eu (at $N=95$) continues the decreasing trend of the
level density with increasing $N$. On the other side of $N=89$, $^{150}$Eu also displays a rather low value. 
 With the points added now at $N=87$ and $N=95$  one can see that the 
low-energy level density of Eu odd-odd nuclei displays  a well defined 
maximum at $N=89$. 

In conclusion, a large number of excited states, close to 60, have been determined 
up to about 1.5 MeV excitation for the
odd-odd nucleus $^{158}$Eu, from a spectrum of the $^{160}$Gd(d,$\alpha$)$^{158}$Eu 
reaction measured at 10$^\circ$. Although the experiment was limited to this 
measurement and could not provide spin/parity value assignments, it allowed 
an examination of the low-energy number of levels in the Eu isotopes with $N$ from 
87 to 95. 
The low-energy level density determined for $^{158}$Eu smoothly continues the decreasing trend of the lighter isotopes. \\

\thanks

   Partial support for this work within the TE67/2018 project with the UEFISCDI 
Romanian research funding Agency is aknowledged. 
The authors thank the technical staff of the Tandem accelerator for the good quality beam. 
This was the last transfer reaction experiment (of the Romanian authors) with the Q3D spectrograph of the tandem accelerator laboratory in Garching -- Munich, which will be closed at the end of 2019. 

\end{document}